\documentclass[amssymb,twocolumn,prd,nofootinbib,showpacs]{revtex4-1}
\pdfoutput=1
\usepackage{graphicx}  % needed for figures
\usepackage{dcolumn}   % needed for some tables}
\usepackage{bm}        % for math
\usepackage{amsmath,amssymb}
\usepackage{hyperref}
\usepackage{color}
\definecolor{purple}{rgb}{0.58,0.0,0.83}
\usepackage{caption}
 \usepackage{soul}
\usepackage[toc,page]{appendix}
\usepackage{bm}        % for math
\usepackage{amsmath,amssymb}
\usepackage{hyperref}
\usepackage{color}
\usepackage{braket}
\definecolor{purple}{rgb}{0.58,0.0,0.83}
\usepackage{caption}
\usepackage[toc,page]{appendix}
\usepackage{subcaption}

\begin{document}

\title{Scalar Field Dark Matter Spectator During Inflation: The Effect of Self-interaction}
\author{Luis E. Padilla}  
\email{epadilla@fis.cinvestav.mx}
\affiliation{Departamento de F\'isica, Centro de Investigaci\'on y de Estudios Avanzados del IPN, 
A.P. 14-740, 07000 M\'exico D.F.,  M\'exico.}
   \author{J. Alberto V\'azquez}  
\email{javazquez@icf.unam.mx}
\affiliation{Instituto de Ciencias F\'isicas, Universidad Nacional Aut\'onoma de M\'exico, Apdo. Postal 48-3, 62251 
Cuernavaca, Morelos, M\'exico}
\author{Tonatiuh Matos}  
\email{tmatos@fis.cinvestav.mx}
\affiliation{Departamento de F\'isica, Centro de Investigaci\'on y de Estudios Avanzados del IPN, A.P. 14-740, 
07000 M\'exico D.F.,  M\'exico.}
   \author{Gabriel Germ\'an}  
\affiliation{Instituto de Ciencias F\'isicas, Universidad Nacional Aut\'onoma de M\'exico, Apdo. Postal 48-3, 62251 
Cuernavaca, Morelos, M\'exico}

\date{\today}

\begin{abstract}
Nowadays cosmological inflation is the most accepted mechanism to explain the  primordial seeds that led to 
the structure formation observed in the Universe. Current observations are in well agreement 
to initial adiabatic conditions, which imply that single-scalar-field inflation may be enough to describe 
the early Universe. However, there are several scenarios where the existence of more than 
a single field could be relevant during this period, for instance, the situation where the so-called 
spectator is present. Within the spectator scenario we can find the possibility that an ultra-light scalar 
field dark matter candidate could coexist with the inflaton. In this work we study this possibility where the additional 
scalar field could be free or self-interacting. We use isocurvature observations to constrain the free parameters of the model.   

\end{abstract}
%\pacs{????}

\maketitle

%%%=========================================================================%%%
\section{Introduction}
\label{introduction}
%%%=========================================================================%%%

It is well accepted that the primordial seeds of the structure formation in the Universe were generated by quantum 
fluctuations given a scalar field (SF) during an inflationary epoch. 
The simplest scenario where density perturbations are carried out by a single inflaton is preferred 
since the initial perturbations are nearly adiabatic \cite{const2,planck,planck2,const3,const4,const5}. 
However the presence of any light field, other than the inflaton, could also fluctuate during inflation and 
contribute to the primordial density perturbations. Of particular interest is the possibility that this extra SF 
can be thought of as the dark matter component (DM). This scenario is usually known 
as the scalar field dark matter (SFDM) and assumes the DM is made of bosonic excitations of an ultra-light SF. 
The typical mass of this field in favor of astrophysical observations is found to be about 
$m\sim 10^{-22}eV/c^2$, which might include self-interaction.

The idea of scalar fields as DM began at the end of the last century (see for example  \cite{Matos:1992qx})
and since then it has been rediscovered 
with different names, for instance Scalar Field Dark Matter \cite{Matos:1998vk},  Fuzzy  DM  \cite{SF3}, 
Wave DM \cite{SF4,SF5}, Bose-Einstein Condensate DM \cite{SF6} or Ultra-light Axion DM \cite{SF7,SF8}, 
amongst many others. Nevertheless, the first systematic study of this hypothesis began in 1998 in \cite{Matos:1998vk}.
The purpose of this SFDM is to resolve the apparent conflict, with observations, that exhibits the
cold dark matter (CDM) formed of weakly-interacting massive particles (WIMPS) \cite{LCDM1,LCDM2}.  
Some of the CDM weaknesses may appear at small-scales within galaxies, e.g. cuspy halo density profiles, overproduction of 
satellite dwarfs within the Local Group and many others, see for instance 
\cite{problem1,problem2,problem3,problem4,problem5}). 
Since then, the SFDM model has been successfully tested by different probes;  
for a review on ultra-light SFDM see \cite{SF9,SF10,SF11,SF12}. 

Despite several works about this model, it continues being unclear the mechanism that generated this ultra-light SFDM particle 
and the time it should have been produced. 
Some works in this direction consider the SFDM as an axion-like particle. 
In such studies the axion is understood as a free field that could be created before or after the inflationary 
era by the misalignment mechanism \cite{laldm}. 
If it is created after inflation then the ultra-light SFDM particle could not account for the total amount of the DM 
because of the production of topological defects, whereas if it is produced during (or before) the inflationary era then 
it is plausible that the SFDM accounts for the total DM in the Universe. 
However, recent studies point out that constrictions coming from cosmological and astrophysical 
observations, in the free field model, are sometimes in tension between them. 
Then it has been postulated the possibility that the SFDM been self-interacting in order to solve such discrepancies. 
Therefore it is natural to question the consequences obtained if the self-interacting candidate 
is generated during the inflationary era. In this work we study such possibility. For simplicity we do not 
contemplate a mechanism of creation for this particle and we only consider the possibility that it was created 
during (or after) the inflationary period with a free or self-interacting potential. It is neccesary to mention that while this work was in progress, in reference \cite{ganaron} it was studied the consecuences of having a spectator SFDM that at the end of inflation its potential remains dominated by the quartic term. However we have to stress-our that in this paper we consider the possible dynamycs that the SFDM could obtain during the inflationary era, obtaining in this way that the work in \cite{ganaron} represents a particular case in our study.    

The paper is organized as follows: First, in section \ref{Generalities} we review the basics of the inflationary scenario. 
Then in section \ref{experimentos} we present the standard observables used to constrain our models. 
Once the mathematical background and the observational constrictions are given, in section \ref{simplest} we 
analyze the possibility that an ultra-light (or self-interacting) SFDM was generated during the inflationary era. 
We provide some limits for the  mass of the scalar field using isocurvature limits and compare them with current  
astrophysical and cosmological observations. 
Then we consider a self-interacting term for the SFDM and we also constrain its free parameters using 
observations. We mention how the self-interacting SFDM model can be helpful to select the inflationary 
potential that causes inflation. Finally in section \ref{conclusions} our conclusions are given.      

%%%=========================================================================%%%
\section{Inflationary scenario: SFDM as a spectator}\label{Generalities}
%%%=========================================================================%%%

This section provides a brief summary of the inflationary process for spectator-like SFs during 
inflation following \cite{twofields} (see also \cite{infm} for a more recent review). Throughout this paper we use natural units ($\hbar=c=1$).

A  SF $\phi_i$ living  during  the  inflationary  era  is  thought
to acquire quantum fluctuations with a primordial power
spectrum measured at Hubble exit as
\begin{equation}\label{eq1}
P_{\phi_i}\simeq\left(\frac{H_*}{2\pi}\right)^2.
\end{equation}
Here  and  after  quantities  with subscript $*$  are evaluated at the Hubble exit. 
Assuming that within the inflationary era there exists several scalar fields,  then it  is  convenient  to  work  on  
a  rotating  basis by defining the \textit{adiabatic field} $\sigma$ parallel  to  the  trajectory  of the field-space 
 and the \textit{entropy fields} $s_i$ perpendicular to the same trajectory. 
 If the background trajectory is a straight-line that evolves in the direction of the field responsible for inflation,
 the scenario is therefore called the \textit{inflaton scenario} (IS). 
 In this  case quantum fluctuations $\delta\sigma$ of  the  adiabatic field  and $\delta s_i$ -- also  
 called  \textit{spectator fields} --,   
 are frozen at Hubble exit\footnote{If the trajectory is curved in the field-space, 
 entropy and adiabatic perturbations are correlated at Hubble exit and the 
 perturbations continue evolving until inflation ends up \cite{twofields}} and start evolving 
 until they re-enter the horizon.
  Considering the case with only one extra SF, 
 additional to the inflaton, then the primordial power spectrum is given by
\begin{equation}\label{eq2}
P_{\sigma^*}(k)=P_{s^*}(k)\simeq\left(\frac{H_*}{2\pi}\right)^2.
\end{equation}

On the other hand, the curvature and isocurvature perturbations are defined as
\begin{equation}\label{RS}
R\equiv\frac{H}{\dot\sigma}\delta \sigma, \ \ \ S=\frac{H}{\dot \sigma}\delta s.
\end{equation}
Then the final power spectra, evaluated at the beginning of the radiation-domination era, are thus
\begin{equation}\label{PRf}
P_R=P_S\simeq P|_*,
\end{equation}
where at linear order in slow-roll parameters
\begin{equation}
P|_*=\frac{1}{2\epsilon}\left(\frac{H_*}{2\pi M_{pl}}\right)^2,
\end{equation}
with $M_{pl}=1.221\times 10^{19}GeV$ being the Planck mass  and 
\[\epsilon\equiv \frac{M_{pl}^2}{2}\left(\frac{V_{,\sigma}}{V}\right)^2.\]
Here $V_{,\sigma}\equiv dV/d\sigma$.
%%%-------------------------------------------------------------------------------------------%%%
\\

\textit{Gravitational waves.-}
%%%-------------------------------------------------------------------------------------------%%%
Given  the  fact  that  scalar  and tensor  perturbations  are  decoupled  at  linear  order,  
the amplitude  of the  gravitational  waves  spectrum and  the tensor-to-scalar ratio $r$, 
in the IS, preserve its values as in the case with no extra spectator fields during the inflationary process.

We  can  see  then  that  the  incorporation  of  these  new fields during the IS will produce 
isocurvature perturbations.  Such perturbations can be used to constrain the free 
parameters of our SFDM models as we will see in the following sections.
%%%=========================================================================%%%
\section{Constraints on inflationary parameters}\label{experimentos}
%%%=========================================================================%%%

In the standard approximation the inflationary observables are given by the tensor-to-scalar ratio $r$, 
the spectral index for adiabatic perturbations $n_R$ and the amplitude for adiabatic perturbations $A_r^2$.   
The constraints of these parameters are quoted at the pivot scale $k_0=0.05 Mpc^{-1}$ 
by \cite{const2,planck,planck2,const3,const4,const5}
\begin{subequations}\label{eq:6}
\begin{equation}\label{amplitude}
A_r^2(k_0)=(2.215^{+0.032}_{-0.079})\times 10^{-9}, \ \ \ \text{at $68\%$ CL},
\end{equation}
\begin{equation}
r_{k_0}<0.064 \ \ \ \text{at $95\%$ CL},
\end{equation}
\begin{equation}\label{n_R}
n_R(k_0)=0.968 \pm 0.006.
\end{equation}
\end{subequations}
Using these measurements we are able to compute the value of the Hubble expansion rate during 
inflation $H_*$ as \cite{H1,H2}
\begin{equation}\label{Hinf}
r = 1.6\times 10^{-5}\left(\frac{H_{*}}{10^{12}GeV}\right)^2.
\end{equation}

As we have noted before, if more than one SF is present during inflation we will obtain isocurvature 
perturbations generated by extra scalar fields perpendicular to the trajectory of the field-space. 
Parameterizing the isocurvature power spectrum for dark matter in terms of the curvature power $P_R(k)$, we have
\begin{equation}\label{isoCDM}
P_{DM}(k) = \frac{\beta_{iso}(k)}{1-\beta_{iso}(k)}P_R(k),
\end{equation}
where $P_{DM}=\braket{\delta \rho_{DM*}/\rho_{DM}}$, $\delta\rho_{DM*}$ are the isocurvature 
perturbations for the DM generated by extra scalar fields during inflation, $\rho_{DM}$ 
is the initial condition of DM and $\beta(k)\equiv P_{DM}/(P_R+P_{DM})$. 
The uncorrelated scale-invariant DM isocurvature is constrained by Planck 
data \cite{const2,const3,const4,const5,planck,planck2} 
at pivot scale $k_0$ as
\begin{equation}\label{betaiso}
\beta_{iso}(k_0)<0.038 \ \ \ \text{at $95\%$ CL}.
\end{equation}
Notice that isocurvature perturbations can be used to compute the inflationary scale, 
just by combining equations \eqref{eq:6} to \eqref{betaiso}.
%
%
%
%%%=========================================================================%%%
\section{CONSTRAINING FREE AND
SELF-INTERACTING ULTRA-LIGHT SFDM
MODELS}\label{simplest}
%%%=========================================================================%%%
In this section we assume the possibility that an ultra-light  SFDM  candidate  coexists  with  the  inflaton  
during the inflationary epoch.  For this, we require that the SFDM candidate be a stable spectator field with negligible 
classical dynamics and energy density. Such scenario is reached by taking the inflaton scenario, i.e. within the  
field space, the evolution of the system is on the inflaton direction $\phi$ whereas the direction perpendicular to the 
trajectory corresponds to the SFDM $\psi$. Notice  this requirement implies that our dark matter candidate evolves 
much slower than the inflaton and its density is smaller than the associated to the inflaton. 

As we mentioned before, to constrain the free parameters of our model the isocurvature perturbations 
have to be taken into account.  For this reason we review the cosmological history that a free 
and a self-interacting scalar field should have gone through the evolution  of  the  Universe,  
to then match the present values of the field with those during the inflationary era and therefore
use the isocurvature constrictions. We also make use of different cosmological  and  astrophysical  
constraints  for each SFDM model in order to compare our results.  
%Let us now explain the way it is obtained. 

First of all let us recap the SFs equations. 
The dynamical evolution of any SF is governed by the Klein-Gordon (KG) equation
\begin{equation}
\Box\Psi-2\frac{dV}{d|\Psi|^2}\Psi=0.
\end{equation}
where $V$ is the potential of the system and $\Box\equiv \nabla_\mu\nabla^\mu$ is the covariant d'Alambert operator. 
When a SF is complex then it is convenient to use the Madelung transformation \cite{madelung}
\begin{equation}
\Psi = \eta \exp[i\theta],
\end{equation}
where $\eta\equiv |\Psi|$ is the magnitude of field $\Psi$ and $\theta$ its phase. 
With this decomposition the 
KG equation splits up on its real and imaginary components as
\begin{subequations}\label{KESFDM}
\begin{equation}\label{KGe1}
\ddot\eta+3H\dot\eta+2\frac{dV}{d|\psi|^2}\eta-\omega^2\eta= 0,
\end{equation}
\begin{equation}\label{KGe2}
\dot\omega \eta + (2\dot\eta+3H\eta)\omega=0,
\end{equation}
\end{subequations}
where $\omega = \dot \theta$ and we have defined $\dot f\equiv df/dt$. Eq. \eqref{KGe2} can be integrated exactly  obtaining 
\begin{equation}
a^3\eta^2\omega=Q,
\end{equation}
where $Q$ is a charge of the SF related to the total number of particles 
\cite{SFphi42,charge1,SFphi41,charge3,charge4}. Plugging 
the last equation into \eqref{KGe1} we obtain that
the radial component of the scalar field follows
\begin{equation}\label{KGe3}
\ddot\eta+3H\dot\eta+M^2\eta-\frac{Q^2}{\eta^3}= 0.
\end{equation}
The term containing $Q$ is given by the complex nature of the SF and is interpreted 
as a ``centrifugal force" \cite{charge4}; $M^2\equiv 2(dV/d|\psi|^2)$ is seen as an effective mass term 
of the scalar field. Notice that if we assume $\Psi$ is the SFDM and we take $Q^2/\eta^3\ll 1$, 
by assuming the SFDM candidate fulfills the slow-roll condition during inflation, 
then the field $\eta$ will remain frozen with value $\eta_i$ until $H\sim M$. Here and for the rest of this 
work subindex $i$ means values right after inflation ends. Then, when $H\sim M$ the field will start 
evolving depending on the effective mass term. In order to get the slow-roll behavior of the SFDM it is 
necessary that $Q\simeq 0$, as explained in \cite{SFphi42}\footnote{In fact the inflationary behavior is an 
attractor solution of the KG equation for a real field in the limit when $M^2\ll H^2$ \cite{atractorinf1,atractorinf2}.
 In this limit the typical dynamics of a real SF is a stiff-like epoch, followed by an inflationary-like era. }. 
 On the other hand if $\Psi$ is the inflaton, $\phi$ it is usually considered as a real field, then in this case $\theta=Q=0$.
 
In this work we consider only situations where the general potential can be 
decomposed as $V(\phi,\psi)=\hat V(\phi)+V_{SFDM}(|\psi|^2)$.

%%%-------------------------------------------------------------------------------------------%%%
%\begin{center}
\subsection{Real ultra-light SFDM candidate}

%\end{center}
%%%-------------------------------------------------------------------------------------------%%%
\subsubsection{Cosmological history} 
The  possibility  that  an  ultra-light  SFDM  candidate could  coexist  with  the  inflaton  
has  been  recently  studied in \cite{SFrev2} considering a potential of the form
\begin{equation}\label{potential_massive}
V(\phi,|\psi|^2)=\hat V(\phi)+\frac{1}{2}m^2\psi^2.
\end{equation}
and by fixing $Q=0$ for the SFDM in Eq. \eqref{KGe3}. Such study was performed by 
assuming the SFDM is an axion-like particle. However their results can be extrapolated for mostly any free SFDM candidate. 
For this particular potential we observe that  $M^2=m^2$. As already mentioned when $H\gg m$ the term with $m^2$ 
in Eq. \eqref{KGe1} can be neglected. Bering in mind the field is slowly rolling during the inflationary era, 
we can neglect second derivatives in \eqref{KGe1}, and thus the field $\psi$ remains frozen with its initial value given by the 
Hubble dragging during the early universe \cite{curvatonatractor}. 
On the other hand, when the $m\sim H$ condition is approached the SFDM starts evolving and oscillates as a free field. 
During its oscillation phase the dependence of $\psi$ respect to the scale factor $a$ is $\psi\sim 1/a^{3/2}$, while its energy density 
behaves as $\rho_{\psi}\sim 1/a^3$  \cite{SFphi41,SFphi42}. So the scalar density of the SFDM can we written as
\begin{equation}\label{rhosfdm}
\rho_\psi = \left\lbrace\begin{array}{ll}
\frac{1}{2}m^2\psi_i^2 & \text{when }H\gg m, \\
\frac{1}{2}m^2\psi_i^2\left(\frac{a_{osc}}{a}\right)^3 & \text{when }H\ll m.
\end{array}\right .
\end{equation}

For typical masses of a SFDM candidate,  $m \sim 10^{-22} eV$, the field started oscillating during the 
radiation-dominated Universe. 
During this period the Hubble parameter evolves in terms of the scale factor as $H\propto a^{-2}$, and the KG 
Eq. \eqref{KGe1} can be solved exactly in terms of $a$. In Ref. \cite{SFrev2}  the initial 
conditions for the free case were obtained by using the evolution of the form \eqref{rhosfdm} and taking into 
account that the entropy of the Universe is conserved. If the total amount of dark matter is made of SFDM 
particles they obtain
\begin{equation}\label{phi_im2}
\psi_i^2\simeq\frac{10^{34}GeV^2}{0.6}\left(\frac{g_{*osc}}{3.36}\right)^{-3/4}\left(\frac{g_{s*osc}}{3.91}\right)\left(\frac{m}{10^{-22}eV}\right)^{-1/2}.
\end{equation}
Here $g_{*osc}$ and $g_{s*osc}$ are the effective degrees of freedom associated to the total particles and to the entropy of the
SFDM oscillations. In particular, for the ultra-light SFDM that started its oscillations during the 
radiation-dominated Universe we have $g_{*osc}= 3.36$ and $g_{s*osc}= 3.91$. 
Notice also that the above expression is not only fulfilled by an ultra-light SFDM, but in fact it is correct for whichever 
SFDM that started its oscillations during the radiation-dominated Universe.

%%%-------------------------------------------------------------------------------------------%%%
\subsubsection{Constraints from isocurvature perturbations}\label{self_const_}
%%%-------------------------------------------------------------------------------------------%%%

For this case we demand the energy density contribution of the SFDM being small during inflation 
(DM dominates right after radiation-matter equality) and hence it is necessary that 
\begin{equation}
\frac{m^2}{2} < \frac{\hat V(\phi)}{\psi_i^2}\simeq \frac{H^2_{*}M_p^2}{\psi_i^2},
\end{equation}
where during inflation our field remains frozen  at value $\psi_i$. 
Notice that for an ultra-light SFDM candidate  ($m\sim 10^{-22}eV$) the above expression is fulfilled for most of the 
initial conditions given by $\psi_i$. 
On the other hand we can constrain isocurvature perturbations generated by a SFDM 
using Eq. \eqref{isoCDM} or equivalently \eqref{betaiso}. 
The analysis was performed in ref \cite{SFrev2} by noticing  that  we  can  re-express  the  primordial 
spectrum as $\delta\rho_\psi/\rho_\psi = 2\delta\psi/\psi_i$ (since  from  Eq.   \eqref{rhosfdm} we have 
$\rho_{SFDM}\propto \psi^2$) which implies that $P_{SFDM}=4P_{\psi}/\psi_i^2$, where $P_\psi$ is given by Eq. \eqref{eq1} 
and $\psi_i$ by Eq. \eqref{phi_im2}, and then compare it with $P_{DM}$ from Eq. \eqref{isoCDM}.  
When such comparison is done they finally obtain the result
\begin{equation}\label{constm}
\frac{m}{10^{-22}\ eV}<\left(\frac{2\times 10^{-4}}{r}\right)^2.
\end{equation}
Then the isocurvature restrictions allow us to constrain the mass parameter of the SFDM in 
terms of the tensor-to-scalar ratio measurements.

The above relation for the mass parameter must be in agreement with  cosmological and astrophysical observations. 
We need to stress out that we cannot use all the constrictions in the literature since some of them consider different 
cosmological evolutions for the SFDM. For example in \cite{SFphi41} it was studied the CMB and the 
Big Bang Nucleosynthesis (BBN) by understanding the SFDM was generated right after inflation with a 
stiff-like equation of state ($p\simeq \rho$). Then, these kind of restrictions are not applicable to our model.    
\\

\textit{Other constraints.-} We start with CMB constrictions. In reference \cite{constn1} the 
CMB was studied in the form of Planck temperature power spectrum, here they obtained  $m\gtrsim 10^{-24}eV$. 
Considering the hydrodynamical representation of the SFDM model, Ref. \cite{constn2} suggests the SFDM's quantum 
pressure as the origin of the offset between dark matter and ordinary matter in Abel 3827. For this purpose they 
required a mass  $m\simeq 2\times 10^{-24}eV$. When the model is tested with the dynamics of dwarf spheroidal 
galaxies (dSphs)  --  Fornax and Sculpture--, in reference \cite{constn3} was obtained a mass constriction 
of $m<0.4\times 10^{-22}eV$ at $97.5\%$. 
The constriction obtained 
when the survival of the cold clump in Ursa and the distribution of globular clusters in Fornax is considered 
requires a mass $m\sim 0.3\ - \ 1\times 10^{-22}eV$ \cite{constn4}. Explaining the half-light mass in the 
ultra-faint dwarfs fits the mass term to be $m\sim 3.7 \ - \ 5.6\times 10^{-22}eV$ \cite{constn5}. 
The model has also been constrained by observations of the  reionization process. In \cite{constn6}, 
using N-body simulations and demanding an ionized fraction of HI of $50\%$ by $z=8$, 
was obtained the result of $m>2.6\times 10^{-23}eV$. Finally, using the Lyman-$\alpha$ 
forest flux power spectrum  demands that the mass parameter fulfills $m\gtrsim 20-30\times 10^{-22}eV$ \cite{constn7,constn8}.

\begin{figure}[h!]
\includegraphics[width=8cm]{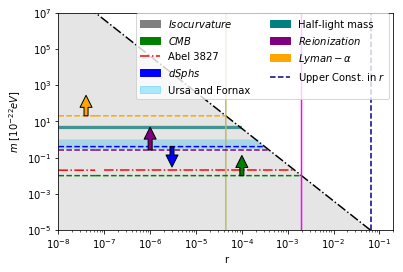}
\caption{Isocurvature constraints for the SFDM candidate.
}\label{constraintsSFDM}
\end{figure}

 Fig. \ref{constraintsSFDM} displays the aforementioned constraints on a $m\ -\ r$ plane. 
 In order to simplify the lecture of the figure we have only plotted the upper (lower) value for the constrictions 
 that fit the mass of the SFDM with an upper (lower) limit. 
 We have also added arrows that points out the region that remains valid for such constrictions. 
 Firstly, the gray region is fulfilled by isocurvature observations \eqref{constm}. 
 The dot-dashed black line corresponds to the equality values in Eq. \eqref{constm}. 
 Then, the figure must be interpreted as follows: suppose we have measured a value for $r$. 
 Notice that such value will intersect with the dot-dashed black line for a given mass $m_{max}$. 
 Then, the masses allowed by the model must be those lower than $m_{max}$. 
 
 The region that fulfills observations obtained by CMB is specified in green, while the value provided by Abel 3827 is given by 
 the dot-dashed red line.  
 The region for dwarf spheroidal galaxies is indicated in blue, Ursa with Fornax in light blue, ultra-faint dwarfs in teal, 
 reionization in purple and Lyman-$\alpha$ in orange. 
 We notice that isocurvature perturbations cannot constrain observations of the dynamics 
 of dSphs galaxies given that both provide an upper limit for the mass of the SFDM. However, 
 the detectability of gravitational waves  and the different constrictions by cosmological and astrophysical observations
 can be used to test the free model. 
 For example, if we ignore by the moment the dynamics of dSphs galaxies and we would like 
 to fulfill at least observations provided by CMB, we should not detect gravitational 
 waves until $r\simeq 1.3\times 10^{-3}$ (fuchsia straight-line), while if we are interested on the rest of observations 
 we should not detect gravitational waves until $r\lesssim 2.33\times 10^{-5}$ (gold straight-line).
 
 These results are important given that \cite{laldm} demonstrated that an ultra-light axion-like dark matter 
 candidate must be present during inflation.  
Then, if $r$ is detected in the near future, it could represent a strong 
constraint for the axion-like particle model. 
 Notice that if we relax the mechanism under this particle is created or if we add an 
 auto-interacting component, we should expect these restrictions be less effective to the model. 

We also plotted the actual upper limit for $r$ in a navy blue dashed-line. By the moment this value is not very 
restrictive for the model since it represents an upper value for $r$. Nevertheless the information it provides is 
that masses smaller than $m_{upper}$ -- the blue dashed-line and black dot-dashed-line intersection --  
 are allowed by the data. However, masses bigger than $m_{upper}$ cannot be discarded 
since the only possible way to do it is if $r$ would be detected.

%%%%-------------------------------------------------------------------------------------------%%%
%\begin{center}
\subsection{Real Self-interacting SFDM Candidate}
\subsubsection{Cosmological history}
%\end{center}0
%%%-------------------------------------------------------------------------------------------%%%

In this section a self-interacting SFDM with a positive interaction is considered. This scenario is described by the general potential
\begin{equation}
V(\phi,|\psi|^2)=\hat V(\phi)+\frac{1}{2}m^2\psi^2+\frac{1}{4}\lambda\psi^4,
\end{equation}
and fixing again $Q=0$ for the SFDM in equation [14]. In what follows we omit the hat in the potential 
$\hat V(\phi)$ in order to make simpler the lecture of the article. Then, always that appears $V$ must be understood that we are 
referring to $\hat V$.

Notice that for this case $M^2=m^2+\lambda\psi^2$. 
As we have previously discussed the effective mass of the field after inflation remains 
constant at $M^2=m^2+\lambda\psi_i^2$ until $M\sim H$. Then, depending of each contribution to $M^2$, we can 
have two different dynamics. 
\\

\begin{figure}[h!]
\includegraphics[width=8cm]{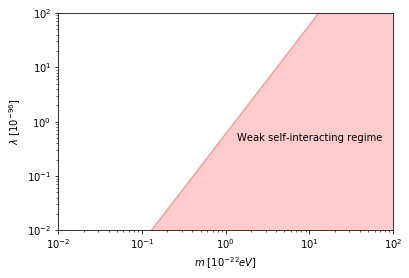}
\caption{Weakly self-interacting regime}
\label{weekregime}
\end{figure}

%%%-------------------------------------------------------------------------------------------%%%
\textbf{\textit{Weakly self-interacting regime.-}} This limit is obtained when the constant term $m^2$ dominates, 
that is when
\begin{equation}\label{consw}
m^2\gg \lambda\psi_i^2.
\end{equation}
In this regime it is possible to ignore the autointeracting term in Eq. \eqref{KGe3} when the oscillations 
of the scalar field begins. 
However, by ignoring this term the field behaves as a free field and from
\eqref{rhosfdm} the field value always decreases. Therefore the autointeracting term never dominates and 
all the cosmological history remains the same as in the pure free SFDM scenario.  In fact, thanks to the 
decreasing behavior of this scenario we can consider that this regime is fulfilled always that 
$m^2\geq \lambda \psi_i^2$ or equivalently when $\lambda\leq m^2/\psi_i^2$. 
If the SFDM oscillations start at the same time than in the free case (which is a good approximation since the effective mass of the SFDM is $M^2 =m^2 +\lambda\psi^2_i \leq 2m^2$), 
we observe from \eqref{phi_im2} that it should be satisfied that
\begin{equation}
\left(\frac{\lambda}{10^{-96}}\right)\leq 0.6\left(\frac{m}{10^{-22}eV}\right)^{5/2}.
\end{equation}
In Fig. \ref{weekregime} we plot the weak limit obtained by our approximation. However this 
overestimates the maximum value of $\lambda$ since the dust-like behavior is obtained when the $\lambda$ 
term is completely negligible.%
\\

%%%-------------------------------------------------------------------------------------------%%%
\textbf{\textit{Strong self-interacting regime.-}} This scenario is obtained when 
\begin{equation}
m^2\ll \lambda\psi_i^2.
\end{equation}
Here the SFDM follows the dynamical equation (see equation \eqref{A5} of appendix [A]) 
\begin{equation}\label{dynamics}
\frac{1}{\psi^2}-\frac{1}{\psi_0^2}= 2\lambda\int_{\phi}^{\phi_0}\frac{d\phi}{V_{,\phi}},
\end{equation}
where subindex $0$ means quantities at the beginning of the inflationary period. Notice that if the inflationary process 
does not last for long enough time --meaning  that $\psi_0^{-2}$ is much smaller than the right hand side value of the 
above expression--, the SFDM remains frozen with value $\psi_0$, while if it lasts  long enough time the SFDM 
reaches an attractor solution.
\\

%%%-------------------------------------------------------------------------------------------%%%
\textit{Attractor behavior of the SF during inflation.-} In the strong self-interacting regime, 
and after enough time, the SFDM follows the 
attractor solution \eqref{atractor0}\footnote{In \cite{curvatonatractor} 
was obtained the attractor behavior for a curvaton-like scalar field in a chaotic-like inflationary scenario. However their results can 
be used as well in this context were the attractor behavior can be easily obtained for whichever inflationary potential.}
\begin{equation}\label{atractor}
\psi_{att} =\left(2\lambda\int_{\phi}^{\phi_0}V^{-1}_{,\phi}d\phi\right)^{-1/2},
\end{equation}
where $\phi_0$ is the value of the inflaton at the beginning of inflation. 
As we can see from the above expression, whether or not the SFDM reaches the attractor solutions 
depends entirely on the inflationary process and the value of the self-interacting term of the SFDM. 
In the above expression we can identify two possible branches:
\begin{itemize}
\item $\psi_{att}<m/\sqrt{\lambda}$

The SFDM follows the attractor solution until $\psi\simeq m/\sqrt{\lambda}$. Then the field 
reaches $\psi_i=m/\sqrt{\lambda}$ for the rest of inflation. Notice that this value corresponds to the upper 
limit that the weakly self-interacting regime allows. 
Then the field starts evolving when $H\sim M\simeq m$ behaving as a free SF. In this way the constrictions 
given in the non-interacting case apply and the initial conditions are also fixed by $\psi_i$. 
Using both relations the $\lambda$ value is approximated 
\begin{equation}\label{wregime}
\left(\frac{\lambda}{10^{-96}}\right)\simeq 0.6\left(\frac{g_{*osc}}{3.36}\right)^{3/4}\left(\frac{g_{s*osc}}{3.91}\right)^{-1}\left(\frac{m}{10^{-22}eV}\right)^{5/2}.
\end{equation}
Without lost of generality the mass term and the auto-interacting constant are rescaled by typical values, that is, 
the mass term is measured in units of $10^{-22}eV$ while the auto-interacting constant in terms of $10^{-96}$.
\\

\item $\psi_{att}>m/\sqrt{\lambda}$

In this scenario the dynamics of the inflaton, given by \eqref{atractor}, implies that the initial condition 
of the field after the inflationary period is
\begin{equation}\label{atractor2}
\psi_{att}^i = \left(2\lambda\int_{\phi_{end}}^{\phi_0}V^{-1}_{,\phi}d\phi\right)^{-1/2},
\end{equation}
where $\phi_{end}$ is the value of the inflaton at the end of inflation. 
We need to stress out that this is the value of the field until its oscillation period starts (i.e. when $M\sim H$).
\end{itemize}

In this scenario and for $\psi_{att}>m/\sqrt{\lambda}$ (regardless of whether the scalar field reached 
the attractor behavior or not) we observe that at the time the SFDM starts its oscillations its effective mass is linear in the field. 
In that regime the scalar field evolves as $\psi\sim 1/a$ and its energy density as $\rho_{\psi}\sim 1/a^4$, 
behaving as radiation. Then, when $m^2 \sim \lambda\psi_t^2$ the effective scalar field mass is now constant, 
obtaining the dust-like behavior already analyzed. 
Therefore, the history of the scalar field density is
\begin{equation}\label{rhosfdmlam}
\rho_\psi = \left\lbrace\begin{array}{ll}
\frac{1}{4}\lambda\psi_i^4 & \text{when }H^2\gg \lambda\psi_i^2 \\
\frac{1}{4}\lambda\psi_i^4\left(\frac{a_{osc}}{a}\right)^4 & \text{when }H_t^2\leq \lambda\psi^2\leq H^2\\
\frac{1}{2}m^2\psi_t^2\left(\frac{a_t}{a}\right)^3 & \text{when } H^2\leq m^2\ \text{and}\ \lambda\psi^2<m^2
\end{array}\right .
\end{equation}
Here sub-index $t$ means quantities measured at transition between radiation-like to dust-like behavior of the SFDM and
\begin{equation}\label{inilamb}
\psi_i^2=\left[\frac{2m^2}{\lambda}\psi_t^2\right]^{1/2}\left(\frac{a_t}{a_{osc}}\right)^2.
\end{equation}
Notice that, for simplicity, we have taken an instantaneous transition between radiation-like to dust-like behaviors. 

Since the auto-interacting KG equation cannot be solved exactly we work with approximated 
solutions. 
By using a pure approximated description of the system,
 \cite{SFphi42} obtained the relation (see its equation 80 and 86 and also \cite{SFphi41})
 \footnote{The reference \cite{SFphi42} obtained this relation by considering a Universe with only a SFDM content. 
 However a similar analysis can be used in a Universe with several types of matter contents.} 
\begin{subequations}
\begin{equation}\label{atoveras}
\left(\frac{a_t}{a_{osc}}\right)^2=\frac{3}{7^{1/3}f^2(\frac{a_s}{r_S})},
\end{equation}
where 
\begin{equation}
f(\sigma)=\frac{1}{s^{1/3}(1+4s)^{1/6}},
\end{equation}
with
\begin{equation}
s=\frac{4\sigma-1+\sqrt{(4\sigma-1)^2+12\sigma}}{6}.
\end{equation}
\end{subequations}
Additionally $r_S=2mG/c^2$ and $a_s=\hbar^2\lambda/4\pi m$. 
Then it follows that $a_s/r_S=\lambda M_{pl}^2/m^2$. 
Rearranging the expression in a more convenient way we have
\begin{equation}
\sigma \simeq 5.93\times 10^{
2}\left(\frac{m}{10^{-22}eV}\right)^{-2}\left(\frac{\lambda}{10^{-96}}\right).
\end{equation}

Notice that when $a_t/a_{osc}\simeq 1$, i.e.  $3/(7^{1/3}f^2(\sigma))\sim 1$, there is no radiation-like epoch. 
This scenario should match with the non-interacting scenario that we present previously.  
Inserting Eq. \eqref{atoveras} into \eqref{inilamb} yields to
\begin{equation}\label{inilamb2}
\psi_i^2=\frac{3}{7^{1/3}f^2(\sigma)}\left[\frac{2m^2}{\lambda}\psi_t^2\right]^{1/2}.
\end{equation}

The relation (\ref{inilamb2}) matches the field at $\psi_t$ with its value right after inflation ends. 
Then if we obtain the value of $\psi_t$ by comparing with quantities at present, with the above expressions 
we can also obtain the value of $\psi_i$. On the other hand, notice that at $a_t$ the scalar field behaves 
as dust with an effective mass $M^2=m^2+\lambda\psi^2_t$. This implies that dust-like oscillations 
of the SF began a little before than in the non-interacting case. 
If we allow $m$ to start its dust-like behavior during the radiation-dominated Universe and 
using the fact that $m^2$ is about the same 
order that $\lambda\psi^2_t$, 
we get that such oscillations start during the same epoch 
than in the non-interacting case. 
In fact because the decreasing behavior of the SF during the dust-like period 
($\psi\sim 1/a^{3/2}$) the auto-interacting term contribution quickly vanishes and then the dynamics of 
the field is described  only by the mass term $m$. Thus, once the dust-like behavior starts, the dynamics 
is described similarly to  the non-interacting case, in such case the condition \eqref{phi_im2} is fulfilled by 
the SF as well, but interchanging subindex $i$ with $t$\footnote{In fact this is a lower limit for the 
strong auto-interacting case.}.
%
%%%=========================================================================%%%
\subsubsection{Constraints from isocurvature perturbations}
%%%=========================================================================%%%

As we have shown in the last section we have two different scenarios for this model: 
a weak self-interacting and a strong self-interacting. In the weak limit our SFDM behaves 
effectively as a free field without auto-interaction, and in such case the constrictions 
for the free field apply to this scenario as well. On the other hand when the auto-interacting 
term is big enough, the SFDM will have a new period with a behavior similar to a radiation-like fluid. 
In this way the constrictions we obtained before will not apply to this model anymore. 

\begin{figure}[h!]
\includegraphics[width=8cm]{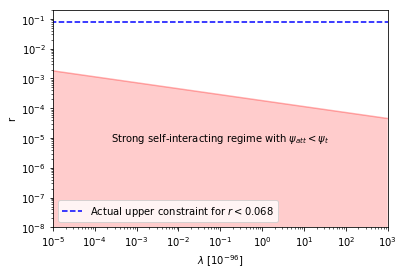}
\caption{Isocurvature constraints for the strong self-interacting regime with $\psi_{att}<\psi_t$.}\label{constraintsSFDMl}
\end{figure}

In the strong self-interacting regime, during the inflationary era, the SFDM follows the solution \eqref{dynamics}.
The value the homogeneous field acquired  after inflation depends on 
the amount of time the inflationary process takes place and then the condition 
$\psi_{att}\lessgtr m/\sqrt{\lambda}\equiv \psi_t$ --if the inflationary period is short, then 
the field $\psi$ remains frozen at value $\psi_i\simeq\psi_0$, while if it lasts long enough the 
SFDM reaches the attractor behavior \eqref{atractor}--. 
If the SFDM reaches the solution \eqref{atractor} and for $\psi_{att}<\psi_t$ the field follows the attractor 
solution until $\psi\simeq \psi_t$. Then the SFDM is frozen at that value and starts oscillating as a free field 
when $m\sim H$. We can constrain this scenario by noticing that it is the same case than the free one 
but with the initial condition $\psi_i=\psi_t$.  Matching Eq. \eqref{phi_im2} with $\psi_t$ and making 
use of the constriction \eqref{constm} we obtain
\begin{equation}\label{lambweek}
\left(\frac{\lambda}{10^{-96}}\right)\leq 0.6\left(\frac{2\times 10^{-4}}{r}\right)^5.
\end{equation}
In Fig. \ref{constraintsSFDMl} we have plotted the above condition that is valid during the strong 
self-interacting regime, when $\psi_{att}<\psi_t$. 
The  pink  region corresponds to the region allowed by isocurvature perturbations in this limit.  
As we  observe the self-interacting term for this model can be constrained in a similar way than the 
mass parameter in the free case. 
This scenario must fulfill the relation \eqref{constm} as well, since its cosmological evolution 
after inflation is only like a free SFDM.

Additionally, in this scenario, the inflationary potential fulfills the condition 
\begin{equation}
\left(\int^{\phi_0}_\phi V_{,\phi}^{-1}d\phi\right)^{-1/2}<\sqrt{2}m.
\end{equation}
We can see that it is very difficult to obtain this relation for an ultra-light SFDM candidate. 
For example, if we consider a chaotic-like inflationary potential, $V(\phi)=\frac{1}{2}M_{inf}^2\phi^2$, 
the above conditions imply that
\begin{equation}\label{chaoticweek}
\left(\log\frac{\phi_0}{\phi}\right)^{-1/2}<\sqrt{2}\frac{m}{M_{inf}}.
\end{equation}
However, for this potential the mass $M_{inf}$ of the inflaton 
that best matches the observations\footnote{This chaotic-like inflationary potential is 
ruled-out now for observations, however we use it as an example in order to obtain 
general constraints for our models.} is of order $M_{inf}\sim 10^{12} GeV$ \cite{Liddle}. 
If now we assume an ultra-light SFDM candidate with a mass $m\sim 10^{-22}eV$, given that the 
cosmological and astrophysical constrictions for this model are the same than in the 
free case and in this scenario it is necessary to obtain an ultra-light SFDM, 
the above conditions imply that the logarithmic part of the expression should be 
lower that $\sim 10^{-43}$. The inflationary behavior for a chaotic-like inflaton ends 
when $\phi_{end}\simeq 2M_{pl}$ \cite{curvatonatractor,Liddle}. Moreover as it is 
explained in \cite{curvatonatractor}, the initial condition of the inflaton cannot be 
arbitrarily large since the stochastic behavior is significant for $\dot\phi H^{-1}<H/2\pi$. 
If the Universe starts when the inflaton escapes from this 
behavior we have that its initial condition should be
\begin{equation}\label{phi_0}
\phi_0\sim 10^5\times M_{pl}\left(\frac{10^{13}GeV}{M_{inf}}\right)^{1/2}.
\end{equation}
where we can easily see that the condition given in \eqref{chaoticweek} cannot be fulfilled. 
In fact if we insert the left part of \eqref{chaoticweek} in equation \eqref{dynamics} and by considering 
the above result and the fact that in this scenario the self-interacting term must be incredible small 
(see equation \eqref{lambweek}) in order to fulfil cosmological and astrophysical constrictions, we can 
observe that the attractor behavior is not reached when it is assumed a chaotic-like inflationary potential.
This implies that if a self-interacting SFDM candidate coexists with the inflaton, 
and that during the begining of the inflationary period it starts in the strong-field scenario, 
it should remain within the strong regime since their conditions are easier to satisfy.

If the SFDM reached the solution \eqref{atractor} and when $\psi_{att}>\psi_t$ we have that the field 
follows the attractor solution during all the inflationary period. 
Hence the initial condition for the SFDM is given by \eqref{atractor2}
\begin{equation}\label{atractor3}
\psi_{att}^i = \left(2\lambda\int_{\phi_{end}}^{\phi_0}V^{-1}_{,\phi}d\phi\right)^{-1/2}.
\end{equation}
Then the SFDM remains frozen at value $\psi_{att}^i$ until $M\sim H$ and starts oscillating with a 
quartic potential. In this scenario the SFDM density behaves as $\rho_{SFDM}\propto \psi^4$ and 
in such case we can write $\delta\rho_{\psi}/\rho_\psi=4\delta\psi/\psi_i$. Therefore the primordial 
isocurvature perturbations for a strong self-interacting SFDM is given by
\begin{equation}
P_{SFDM}(k)=\left(\frac{2H_*}{\pi\psi_i}\right)^2.
\end{equation}
In the last section we showed the relation of the initial condition  with the value of the 
field today. Using eqs. \eqref{inilamb2} and \eqref{phi_im2}, 
with $g_{*osc}=3.36$ and $g_{s*osc}=3.91$  and appropriate units we obtain
\begin{equation}\label{constr4}
r<\frac{1.172\times 10^{-4}}{7^{1/3}f^2(\sigma)}\left[\frac{2
\left(\frac{m}{10^{-22}eV}\right)^{3/2}}{\left(\frac{\lambda}{10^{-96}}\right)}\right]^{1/2}.
\end{equation}
Notice that the above relation is independent of whether the SFDM followed the attractor 
solution or not and therefore the result is general, always that the SFDM remains in the strong self-interacting regime 
at the end of inflation and its dust-like behavior started in the radiation dominated Universe. 

Similarly to the free case, the above relation must be compared with observations in order to 
get constrictions for  the strong self-interacting SFDM scenario.  
\\

\textit{Other constraints}.- In \cite{constl1} it was studied the possibility that a SFDM candidate could be self-interacting. 
In their work they constrain the ratio $\Lambda\equiv m/\lambda^{1/4}$ to be $\Lambda\sim 1eV$ by analyzing the 
line-of-sight velocity dispersion for the eight dSphs satellites of the Milky Way (MW). This study was complementary to the ones done in \cite{SF6,constl3,constl4,constl5} where the SFDM model was studied by using rotational curves of the 
most Dark Matter dominated galaxies from different surveys and where they obtained the 
constriction $\Lambda\sim 2.6-2.9eV$. On the other hand, in \cite{constl6} the model was studied in a 
cosmological context by demanding that the SFDM candidate behaves as a dust-like component before the time of matter-radiation equality. In such work they obtain the result $\Lambda>0.8eV$. Finally, it was also possible to test the self-interacting scenario by considering the number of extra relativistic species at BBN. When such observations are confronted with the SFDM it is obtained that it must be fulfilled that $\Lambda\gtrsim 4eV$ (see \cite{constl3}).    

\begin{figure}[h!]
\includegraphics[width=8cm]{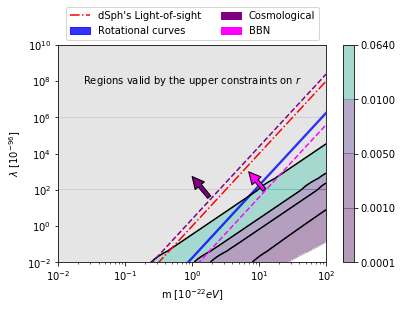}
\caption{Isocurvature constraints for the strong self-interacting scenario with $\psi_{att}>\psi_t$.
}\label{constraintsSFDMls}
\end{figure}

In Fig. \ref{constraintsSFDMls} we have plotted contour levels of the numerical value of the right 
hand side of the relation \eqref{constr4}. 
The grey region corresponds to values larger than $0.064$ which is the actual upper constraint on 
tensor-to-scalar ratio. 
This means that within that region we are certain that \eqref{constr4} is fulfilled and, by the moment, 
just this region is completely allowed by observations. With this in mind the plot should be understood as follows: 
let us suppose that in the near future we measure a value for $r$. Such value will coincide with a curve in 
Fig. \ref{constraintsSFDMls}. Then, the parameter space  allowed by data should 
be the one where the contour levels are bigger than the detected value of $r$, while the one with smaller 
values in the contour levels must be discarded. Similar to the free case, if $r$ is not detected but it 
continues  with an upper limit, this implies that regions with contour levels bigger than 
such upper value will be allowed by the data, however regions with smaller values can not be discarded until $r$ is detected.

The  $\Lambda$ value that satisfies observations of dSph's line-of-sight is given by the dot-dashed red line, 
while the region of parameters necessary for rotational curves is presented in blue. Similar 
to the plot for the free case (fig. \ref{constraintsSFDM}) we plotted the cosmological constrictions in purple by 
drawing a curve that refers to the upper value of $\Lambda$ and an arrow that points out to the valid region 
 from the constriction. We did the same for the observations for BBN but in color fuchsia. The 
white region corresponds to the weak limit.  
We can see from Fig. \ref{constraintsSFDMls} that it is possible to fulfill observations for whichever 
value for the mass as long as the self-interacting constant is large enough. In other words, for a given 
mass, the measurement of $r$ can only constrain the self-interacting constant with a lower limit. 
Isocurvature observations (or equivalently observations on $r$) can also help to impose 
 upper values from other kind of constrictions. As an example let us suppose that we want to be completely 
sure that the measurements obtained by rotational curves are fulfilled. Then, given the actual 
upper constraints in $r$, the only region of parameters that we can be sure that fulfills 
observations are the ones in the blue region that are also inside the gray region. 

Remark: This scenario is of special interest given that it is natural to avoid isocurvature perturbations 
when the auto-interacting term of the SF is big enough. 
Additionally, if the SFDM reached the attractor solution it is possible to justify the initial conditions for the SFDM model.

Similarly to the above description we can compute general constraints for the inflationary 
potential that should generate inflation on these kind of scenarios. First we have that
\begin{equation}
\left(\int^{\phi_0}_\phi V_{,\phi}^{-1}d\phi\right)^{-1/2}>\sqrt{2}m,
\end{equation}
which is very easy to fulfill as we saw in the chaotic-like example. Using isocurvature constriction we also have
\begin{equation}
r<\frac{0.6\times 10^{40}}{\left(\frac{\lambda}{10^{-96}}\right)}\left(\frac{1}{\int_{\phi_{end}}^{\phi_0}V_{,\phi}^{-1}d\phi}\right),
\end{equation}
that can be satisfied as far as the auto-interacting term and the integral are small enough; 
for example in the chaotic-like scenario by using \eqref{chaoticweek} and \eqref{phi_0}
and taking $M_{inf}\simeq 10^{-6}M_{pl} $, we can obtain the constriction 
\begin{equation}
\left(\frac{\lambda}{10^{-96}}\right)<\frac{0.3288\times 10^{84}}{r}.
\end{equation}
that is easily satisfied for whichever value of $\lambda$ of our interest.
If now we compare \eqref{atractor2} and \eqref{inilamb2} we have
\begin{equation}\label{inilamb3}
\left(\int_{\phi_{end}}^{\phi_0}V^{-1}_{,\phi}d\phi\right)^{-1}=\frac{6}{7^{1/3}f^2(\sigma)}\left(2m^2\lambda\psi_t^2\right)^{1/2}.
\end{equation}
This relation is interpreted as follows: consider that the auto-interacting SFDM candidate 
coexists with the inflaton, and it reached the attractor solution \eqref{atractor}, 
and suppose there are several measurements constraining the mass 
parameter $m$ as well as the auto-interacting parameter $\lambda$, therefore such constraints are translated into 
restrictions to the inflationary potential. In order to use the above expression to constraint inflationary potentials, 
we need to be sure that the term $\psi_0^{-2}$ can be ignored in \eqref{dynamics} i.e. the right-hand side term 
in equation \eqref{dynamics} is big enough compared with $\psi_0^{-2}$. 

It is also necessary to be careful that the SFDM does not come to dominate the inflationary period. 
This is guarantee by demanding that
\begin{equation}
\frac{\lambda}{4} < \frac{H_*^2 M_{pl}^2}{\psi_i^4},
\end{equation}
or in terms of \eqref{atractor3}
\begin{equation}\label{constd}
\lambda>\left(16H_*^2M_{pl}^2\left(\int_{\phi_{end}}^{\phi_0}V_{,\phi}^{-1}d\phi\right)^2\right)^{-1}.
\end{equation}
Taking the chaotic-like example and using $H_*= 10^{14} GeV$, we obtain the constriction
\begin{equation}\label{lowerlambda}
\frac{\lambda}{10^{-96}} > 1.444\times 10^{77}.
\end{equation}
Notice that the above expression requires $\lambda>10^{-19}$ and then by considering the cosmological 
and astrophysical constrictions ($\Lambda\sim O(1)eV$) we should have a mass parameter of order $m\sim 10^{-5}eV$. 
We have to stress out that \eqref{lowerlambda} is obtained for a  chaotic-like potential, then, 
depending on the inflationary potential we will have different limits allowed for the self-interacting scenario. 
On the other hand if Eq. \eqref{constd} is not fulfilled and we remain in the strong-self interacting 
model it should be necessary to consider a two-field inflationary scenario where the SFDM could obtain a 
non-negligible dynamics during inflation.

%%%-------------------------------------------------------------------------------------------%%%
\subsection{Complex SFDM generalization}
%%%-------------------------------------------------------------------------------------------%%%

As we have seen at the beginning of this section, when we consider a complex scalar field its dynamics is modified 
only by the centrifugal term (see eq. \eqref{KGe3}). However as it was mentioned in \cite{SFphi42} 
such term does not affect the dynamics of the field at cosmological levels, obtaining then that a 
complex scalar field and a real scalar field have the same cosmological history in the Universe. 
In this way if we take that our complex SFDM fulfills slow-roll conditions during inflation then its 
constrictions for isocurvature perturbations must be the same than in the real field analogue. \\

%%%-------------------------------------------------------------------------------------------%%%
\section{Discusions and conclusions}\label{conclusions}
%%%-------------------------------------------------------------------------------------------%%%

In this paper we have studied the possibility that a free or a self-interacting SFDM particle could coexist with the 
inflaton during inflation. In our assumptions we have considered the SFDM as a spectator in the inflationary process. 
Then, the SFDM contributes to the primordial spectrum by generating isocurvature perturbations. 
By using the actual upper constraints in the measurements of the tensor-to-scalar ratio $r$ 
was possible to test the free parameters for each model. As we discussed, at the moment it is difficult 
to rule-out some regions of parameters, however it could be possible if $r$ is measured soon.   

Our main results are shown in figures \ref{constraintsSFDM} and \ref{constraintsSFDMls}. 
In Fig. \ref{constraintsSFDM} we have identified the masses allowed in the free model by 
isocurvature as well as cosmological and astrophysical observations. We obtained that in order to 
fulfill the constrictions imposed by CMB we should not detect gravitational waves 
until $r\simeq 1.3\times 10^{-3}$, while if we were interested in fulfilling all the observations, 
we should not detect gravitational waves until $r\lesssim 2.33\times 10^{-5}$. This last result is 
important given that the detectability of gravitational waves could represent a strong constriction for 
the free model. Analogously, in figure  \ref{constraintsSFDMls} we have plotted in a $m\ -\ \lambda$ 
plane the region of parameters for the strong-self-interacting model that are allowed by observations. 
We noticed that for a given mass of the SFDM it is always possible to avoid isocurvature 
constrictions and fit astrophysical and cosmological observations if a large enough self-interaction is added. 
Then we notice with this result that the addition of a self-interacting component to the SFDM 
seems to be a natural solution for the model given that is possible to fulfill naturally 
all the constrictions that the model has. On the other hand we explain how the SFDM spectator 
scenario could help to choose the inflationary potential responsible to produce the inflationary period.

\section{Acknowledgments}
We would like to thank Abril S\'uarez for discussions on this subject. This work was partially 
supported by CONACyT M\'exico under grants CB-2014-01 No. 240512, Project
No. 269652 and Fronteras Project 281;
Xiuhcoatl and Abacus clusters at Cinvestav, IPN; I0101/131/07 C-234/07 of the Instituto
Avanzado de Cosmolog\'ia (IAC) collaboration (http://www.iac.edu.mx/). TM is partially supported by Conacyt through the 
Fondo Sectorial de Investigaci\'on para la Educaci\'on, grant CB-2014-1, No. 240512.
JAV acknowledges the support provided by FOSEC SEP-CONACYT Investigaci\'on B\'asica 
A1-S-21925, and UNAM-PAPIIT  IA102219. LEP was supported by CONACyT M\'exico.

\appendix
\section{The attractor behaviour for the SFDM candidate}

In this appendix we comment about the attractor behaviour of the strong-self-interacting SFDM during inflation. 
For this purpose let us remember the dynamical equations that the Universe follows when it contains only two 
real scalar fields $\phi$ and $\psi$. In that case the Universe is described by the Friedmann and Klein-Gordon differential equations
\begin{subequations}
\begin{equation}
H^2 = \frac{1}{3M_{pl}^2}\left[\frac{1}{2}\dot\phi+\frac{1}{2}\dot\psi+V(\phi,\psi)\right],
\end{equation}
\begin{equation}\label{KGfull}
\ddot\phi_i+3H\dot\phi_i+V_{,\phi_i}=0
\end{equation}
\end{subequations} 
where $\dot\phi_i\equiv d\phi_i/dt$, $V_{,\phi_i}\equiv dV/d\phi_i$, $M_{pl}=1.221\times 10^{19} GeV$ is 
the Planck mass and $\phi_{1,2}=\phi,\psi$. In what follows we consider the full potential 
$V(\phi,\psi)\simeq \hat V(\phi)+m^2\psi^2/2+\lambda\psi^4/4$. 

In the IS it is assumed that the Universe is dominated by the inflaton and that it is slowly-rolling during that process, 
i.e. that the slow-roll parameters  
\begin{subequations}
\begin{equation}\label{slowroll}
\epsilon_\phi\equiv \frac{M_{pl}^2}{2}\left(\frac{V_{,\phi}}{V}\right)^2, \ \ \ \ \eta_{\phi}\equiv M_{pl}^2\left(\frac{V_{,\phi\phi}}{V}\right)
\end{equation}
are small ($\epsilon_\phi\ll 1$ and $\eta_\phi\ll 1$). In that case the Friedmann equation and the Klein-Gordon 
equation associated to the inflaton are reduced to
\begin{equation}\label{slowhubble}
H^2\simeq \frac{\hat V(\phi)}{3M_{pl}^2},
\end{equation}
\begin{equation}\label{kg-inflaton}
3H\dot\phi+V_{,\phi}=0,
\end{equation}
\end{subequations}
while the dynamics for the SFDM continue being, in general, described by equation \eqref{KGfull}.

%%=========================================================================
\subsection{Justifying the slow-roll condition for the SFDM candidate}
%%%-------------------------------------------------------------------------------------------%%%

In order to obtain a slowling-rolling SFDM during the inflationary process it is necessary that it fulfills a similar relation 
that the one  by the inflaton, i.e. the slow-roll parameters associated for the SFDM $\epsilon_\psi$ and $\eta_\psi$ 
--defined in a similar way than in \eqref{slowroll}-- being small ($\epsilon_\psi,\eta_\psi\ll 1$). 
Considering that in the IS $V\simeq \hat V$ and from \eqref{slowhubble} $\hat V\simeq 3M_{pl}^2H^2$ we obtain
the slow-roll parameters for the SFDM can be written as
\begin{equation}\label{sr_}
\epsilon_\psi\simeq \frac{1}{2}\left(\frac{\lambda\psi^3}{3M_{pl}H_*^2}\right)^2, \ \ \ \ \eta_\psi\simeq  \frac{\lambda\psi^2}{H_*^2}.
\end{equation}
where, as mentioned before, $H_*$ is the hubble parameter measured at the epoch of inflation and we 
have considered that we are in the strong-self-interacting regime. Typically the self-interacting 
scenario is constrained by the ratio $m/\lambda^{1/4}\sim O(1)eV$ (see subsection \ref{self_const_} 
for more detail). As an example, notice that if we consider an ultra-light SFDM candidate with a mass 
of order $10^{-22} eV$ then the self-interaction should be extremely small $\lambda\sim 10^{-88}$. 
In that case we can observe that the slow-roll condition can be fulfilled for most of the values of the 
field $\psi$. However notice that the constraints in this self-interacting scenario allows the SFDM to 
obtain bigger masses with bigger self-interaction. Then our study is correct always that the slow-roll 
parameters \eqref{sr_} are small during inflation.
\subsection{Attractor solution for the SFDM}

The dynamic of the SFDM during inflation is described by equation \eqref{kg-inflaton} but 
interchanging $\phi$ for $\psi$. In that case notice that both fields must follows the relation
\begin{equation}
\frac{d\psi}{V_{,\psi}}=\frac{d\phi}{V_{,\phi}}
\end{equation}
In the strong-self-interacting regime (where $V_{,\psi}\simeq \lambda\psi^3$) the above equation results in 
\begin{equation}\label{A5}
\frac{1}{\psi^2}-\frac{1}{\psi_0^2}= 2\lambda\int_{\phi}^{\phi_0}\frac{d\phi}{V_{,\phi}}.
\end{equation}
Then, after enough time $\psi$ becomes far smaller than $\psi_0$ and then the field reaches the attractor solution
\begin{equation}\label{atractor0}
\psi_{att} =\left(2\lambda\int_{\phi}^{\phi_0}V^{-1}_{,\phi}d\phi\right)^{-1/2},
\end{equation}
Notice that the time needed to obtain the attractor behavior for the SFDM is described by the inflationary 
potential and the self-interaction parameter. Then, the attractor behavior is reached more quickly for the 
SFDM models with large self-interaction compared to models
with small self-interaction. As an example notice that when the SFDM has an extremely small self-interaction 
and if the inflationary period does not last for a long time, the attractor behavior is not reached and then 
the dynamics is described by \eqref{A5}. In fact if the self-interaction is extremely small we can 
approximate $\psi_{end}\sim \psi_0$, where $\psi_{end}$ is the value of $\psi$ at the end of inflation.

\end{document}